\documentclass[10pt,aps,prl,twocolumn,showpacs,reprint]{revtex4-1}
\usepackage{amsmath,amssymb,graphicx,color}
\usepackage{hyperref}

\definecolor{link_color}{rgb}{0.2,0.2,0.6}

\hypersetup{colorlinks = true, linkcolor = link_color, citecolor = link_color, urlcolor = link_color}

\newcommand\smallsymb[1]{\mbox{\scalebox{0.46}{$#1$}}}

\newcommand\mediumsymb[1]{\mbox{\scalebox{0.69}{$#1$}}}
\newcommand\symb[1]{\mbox{\scalebox{0.55}{$#1$}}}
\newcommand*\Laplace{\mathop{}\!\mathbin\bigtriangleup}
\newcommand*\shortminus{\scalebox{0.5}[1.0]{$-$}}

\begin{document}
\title{Necessity of Time-Reversal Symmetry Breaking for the Polar Kerr Effect in Linear Response}

\author{Weejee Cho}
\author{Steven A. Kivelson}

\affiliation{Department of Physics, Stanford University, Stanford, California 94305, USA}
\date{\today}
\begin{abstract}
We show that, measured in a backscattering geometry, the polar Kerr effect is absent if the nonlocal electromagnetic response function respects Onsager symmetry, characteristic of thermodynamic states that preserve time-reversal symmetry. A key element is an expression for the reflectivity tensor in terms of the retarded Green's function.\end{abstract}
\maketitle

The polar Kerr effect \cite{kerr77} refers to rotation of the polarization of light upon reflection due to a component of the magnetization perpendicular to the reflecting surface. More generally, it is interpreted as a measure of the corresponding pattern of broken symmetry. Inevitably, a sharp onset of Kerr rotation indicates a symmetry-breaking phase transition. Observation of this phenomenon has, indeed, played a crucial role in identifying broken time-reversal symmetry in unconventional superconducting states, e.g., of Sr${}_{2}$RuO${}_{4}$ \cite{xia_srru}, UPt${}_{3}$ \cite{14schemm_science}, and URu${}_{2}$Si${}_{2}$ \cite{15schemm_prb}. Kerr onsets have also been found in the pseudogap regime of the cuprates \cite{xia_ybco, he11science, karapetyan12prl, karapetyan14prl}, although they are somewhat rounded and their interpretation less clear.

In Refs. \cite{xia_srru, 14schemm_science, 15schemm_prb, xia_ybco, he11science, karapetyan12prl, karapetyan14prl}, the polar Kerr effect was measured in a backscattering geometry. In the linear response regime, the corresponding Kerr rotation originates from the ac Hall conductivity in the dimensions perpendicular to the wave vector of light \cite{argyres55pr,wang14prb}, provided that spatial dispersion (nonlocality) is negligible \cite{ignore_nonlocal}. For this Hall conductivity to be nonzero, the medium must break time-reversal symmetry and mirror symmetries about all planes parallel to the wave vector \cite{goryo08prb}.

However, given the extremely high sensitivity of the Kerr measurements ($\lesssim 100\,\mathrm{nrad}$), it is conceivable that a detectable Kerr rotation arises entirely due to spatial dispersion effects in a medium with zero Hall conductivity. An important question is whether the symmetry requirements discussed in the previous paragraph carry over even if spatial dispersion is taken into account. It is easy to see that the mirror symmetries must be broken in any case. The sense of Kerr rotation is reversed in a mirror parallel to the wave vector, so if the system is invariant under this mirror reflection, the Kerr angle must vanish. Consequences of time-reversal symmetry are more subtle. Kerr rotation is typically described by the macroscopic Maxwell equations which are not invariant under time reversal in the presence of dissipation, even if there is no broken time-reversal symmetry per se.

Indeed, there have been proposals \cite{arfi92prb, bungay93prb} that the polar Kerr effect in backscattering can result from natural optical activity \cite{note0}, a spatial dispersion effect, even if time-reversal symmetry is unbroken. In a medium with natural optical activity, the speed and damping of circularly polarized light depend on the handedness, and it is plausible that this gives rise to nonzero Kerr rotation. In Refs. \cite{arfi92prb, bungay93prb}, the authors computed the Kerr angle in the long-wavelength limit, where spatial dispersion is treated to first order in the wave vector, and obtained nonzero results. Several researchers \cite{mineev2010, mineev2013, hosur2013, yakovenko13prl} recently adopted this idea to interpret Kerr signals from unconventional superconductors.

Yet there were earlier studies suggesting the contrary \cite{agranovich1973phenomenological,schlagheck1975boundary}. They noted that the macroscopic Maxwell equations should be consistent with the Onsager symmetry \cite{callen1} of the electromagnetic response function. This is a consequence of \emph{time-reversal symmetry} and \emph{thermal equilibrium}, and holds whether the medium is dissipative or not. In the long-wavelength limit, this consideration leads to electromagnetic boundary conditions (at the boundary of the reflecting medium) different from the ones used in the studies that found nonzero Kerr rotation \cite{arfi92prb, bungay93prb}. The Kerr angle computed with these modified boundary conditions vanishes. Moreover, there were arguments that did not rely on the long-wavelength approximation but reached the same conclusion \cite{halperin1992,shelankov92}.

The significance of Onsager symmetry has been re-emphasized in a number of recent papers \cite{armitage2014jones, fried2014kerr,kapitulnik2015notes,yakovenko2015tilted}. This has led to the retraction \cite{mineev2014erratum,hosur2014revisited,yakovenko14erratum} of the proposals that various measured Kerr signals are due to optical activity alone.

Still, as far as we know, the consequences of Onsager symmetry have not been fully clarified. One should be able to see how it constrains the reflectivity tensor in such a way that Kerr rotation is forbidden. However, existing results, if not relying on the long-wavelength limit, either bypass dealing with the reflectivity tensor \cite{halperin1992, fried2014kerr} or take the principle of optical reciprocity \cite{perrin1942polarization,dehoop1960reciprocity} as the fundamental postulate \cite{shelankov92,armitage2014jones}, but this principle really should be derived from the Onsager symmetry of the response functions. The main obstacle is that without an approximation, there is no obvious way to express the reflectivity tensor in terms of the (nonlocal) response function of the scattering medium. In this work, we avoid this difficulty by expressing the reflectivity tensor in terms of the retarded Green's function of the electromagnetic wave equation. Then, it can be shown that the symmetry of the response function is inherited by the Green's function and, hence, by the reflectivity tensor. From this, we demonstrate, in the framework of nonlocal electrodynamics, that Onsager symmetry leads to optical reciprocity in reflection \cite{carminati1998reciprocity} and, as a special case, the absence of the polar Kerr effect in backscattering.

\begin{figure}[t]
\centering
\includegraphics[scale=0.88]{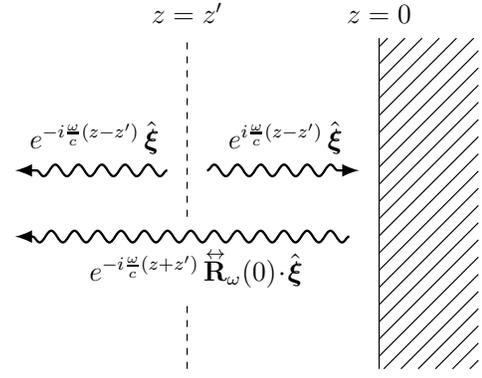}\\
\caption{An example illustrating the relation between the Green's function and the reflectivity tensor: The source $\textbf{J}(\textbf{r}) \propto \delta(z-z^{\prime})\,\hat{\boldsymbol{\xi}}$, where $\hat{\boldsymbol{\xi}}\perp \hat{\textbf{z}}$, generates a pair of counter-propagating plane waves. One of them is reflected from the medium.}
\label{fig}
\end{figure}

We begin by considering the problem of light reflection from an arbitrary linear scattering medium. Without loss of generality, we assume that the medium is entirely in the right half-space ($z>0$), as in the example illustrated by Fig. \ref{fig}. Suppose that a right-moving plane wave of frequency $\omega$ and transverse wave vector $\textbf{k}_{\scriptscriptstyle{\parallel}}^{\prime} \equiv (k_{x}^{\prime},k_{y}^{\prime})$ is incident on the medium. This wave is of the form $e^{i\textbf{k}_{\symb{+}}^{\prime}\cdot \textbf{r} }\,\boldsymbol{\mathcal{E}}_{\!\scriptscriptstyle{\perp}}$. (A harmonic time dependence $e^{\shortminus i\omega t}$ is assumed throughout.) Here, $\textbf{k}_{\mediumsymb{+}}^{\prime}\equiv (\textbf{k}_{\scriptscriptstyle{\parallel}}^{\prime},  +k_{z}^{\prime})$ with \cite{note1}
\begin{equation}
\label{kzdef}
k_{z}^{\prime} \equiv
\begin{cases}
\mathrm{sgn}(\omega)\sqrt{\omega^{2}/c^{2} - \textbf{k}_{\scriptscriptstyle{\parallel}}^{\prime\,2}} &\big(|\textbf{k}_{\scriptscriptstyle{\parallel}}^{\prime}|<|\omega|/c\big),\\\\
i\sqrt{\textbf{k}_{\scriptscriptstyle{\parallel}}^{\prime\,2}-\omega^{2}/c^{2}} &\big(|\textbf{k}_{\scriptscriptstyle{\parallel}}^{\prime}|\ge|\omega|/c\big),
\end{cases}
\end{equation}
where $c$ is the speed of light in vacuum; $\boldsymbol{\mathcal{E}}_{\!\scriptscriptstyle{\perp}}$ can be any vector satisfying $\textbf{k}_{\mediumsymb{+}}^{\prime}\cdot\boldsymbol{\mathcal{E}}_{\!\scriptscriptstyle{\perp}}=0$. One should seek a solution to the electromagnetic wave equation with the following properties: for $z<0$, the solution is the sum of the incident wave $e^{i\textbf{k}_{\symb{+}}^{\prime}\cdot \textbf{r} }\,\boldsymbol{\mathcal{E}}_{\!\scriptscriptstyle{\perp}}$ and a left-moving term corresponding to the reflected wave; for $z>0$, it represents the transmitted wave.

The wave equation is linear, and so is the relationship between the incident and reflected waves. The general form of a reflected (left-moving) wave linearly related to $e^{i\textbf{k}_{\symb{+}}^{\prime}\cdot \textbf{r} }\,\boldsymbol{\mathcal{E}}_{\!\scriptscriptstyle{\perp}}$ is
\begin{equation}
\label{reflected}
\int \frac{d^{2}\textbf{k}_{\scriptscriptstyle{\parallel}}}{(2\pi)^{2}} \,e^{i\textbf{k}_{\symb{-}}\cdot \textbf{r}}\,\tensor{\textbf{R}}_{\omega}(\textbf{k}_{\scriptscriptstyle{\parallel}},\textbf{k}_{\scriptscriptstyle{\parallel}}^{\prime}) \cdot\boldsymbol{\mathcal{E}}_{\!\scriptscriptstyle{\perp}} \quad \ (z<0).
\end{equation}
Here, $\textbf{k}_{\mediumsymb{-}}\equiv(\textbf{k}_{\scriptscriptstyle{\parallel}}, -k_{z})$, and $k_{z}$ is defined analogously to $k_{z}^{\prime}$. The  \emph{reflectivity tensor} $\tensor{\textbf{R}}_{\omega}(\textbf{k}_{\scriptscriptstyle{\parallel}},\textbf{k}_{\scriptscriptstyle{\parallel}}^{\prime})$ satisfies the transversality conditions $\textbf{k}_{\mediumsymb{-}}\cdot\tensor{\textbf{R}}_{\omega}(\textbf{k}_{\scriptscriptstyle{\parallel}},\textbf{k}_{\scriptscriptstyle{\parallel}}^{\prime})
=\tensor{\textbf{R}}_{\omega}(\textbf{k}_{\scriptscriptstyle{\parallel}},\textbf{k}_{\scriptscriptstyle{\parallel}}^{\prime})\cdot\textbf{k}_{\mediumsymb{+}}^{\prime}=0$.

The backscattered component of the reflected wave is given by the integrand of Eq. (\ref{reflected}) at $\textbf{k}_{\scriptscriptstyle{\parallel}} = -\textbf{k}_{\scriptscriptstyle{\parallel}}^{\prime}$. Its polarization can be rotated relative to that of the incident wave only if the antisymmetric part of $\tensor{\textbf{R}}_{\omega}(-\textbf{k}_{\scriptscriptstyle{\parallel}}^{\prime},\textbf{k}_{\scriptscriptstyle{\parallel}}^{\prime})$ does not vanish. That is, Kerr rotation is absent when
\begin{equation}
\label{kerrabsence}
\tensor{\textbf{R}}_{\omega}(-\textbf{k}_{\scriptscriptstyle{\parallel}}^{\prime},\textbf{k}_{\scriptscriptstyle{\parallel}}^{\prime}) = \big[\tensor{\textbf{R}}_{\omega}(-\textbf{k}_{\scriptscriptstyle{\parallel}}^{\prime},\textbf{k}_{\scriptscriptstyle{\parallel}}^{\prime})\big]^{T}.
\end{equation}
If the medium is uniform in $x$ and $y$, the reflectivity tensor has the form
\begin{equation}
\tensor{\textbf{R}}_{\omega}(\textbf{k}_{\scriptscriptstyle{\parallel}},\textbf{k}_{\scriptscriptstyle{\parallel}}^{\prime}) = (2\pi)^{2}\delta^{(2)} (\textbf{k}_{\scriptscriptstyle{\parallel}}-\textbf{k}_{\scriptscriptstyle{\parallel}}^{\prime})\,\tensor{\textbf{R}}_{\omega}(\textbf{k}_{\scriptscriptstyle{\parallel}}^{\prime}).
\end{equation}
In this case, Eq. (\ref{kerrabsence}) reduces to a simpler condition $\tensor{\textbf{R}}_{\omega}(0) = \big[\tensor{\textbf{R}}_{\omega}(0)\big]^{T}$.

The next step is to relate light reflection to the retarded Green's function in the frequency domain, denoted by $\tensor{\textbf{G}}_{\omega}(\textbf{r},\textbf{r}^{\prime})$. It is a tensor such that $\tensor{\textbf{G}}_{\omega}(\textbf{r},\textbf{r}^{\prime})\cdot\hat{\boldsymbol{\xi}}$ is the electric field at $\textbf{r}$ due to a harmonic point source $i\omega \mu_{0} \textbf{J}(\textbf{r}) = \delta^{(3)}(\textbf{r}-\textbf{r}^{\prime})\hat{\boldsymbol{\xi}}$, where $\hat{\boldsymbol{\xi}}$ is an arbitrary unit vector. In the presence of a medium, $\tensor{\textbf{G}}_{\omega}$ also includes the effect of scattering. One can decompose the electric field into the incident wave $\tensor{\textbf{G}}_{\omega}^{0}(\textbf{r},\textbf{r}^{\prime})\cdot \hat{\boldsymbol{\xi}}$ and the reflected wave $\big[\tensor{\textbf{G}}_{\omega}(\textbf{r},\textbf{r}^{\prime}) - \tensor{\textbf{G}}_{\omega}^{0}(\textbf{r},\textbf{r}^{\prime})\big]\cdot \hat{\boldsymbol{\xi}}$, where $\tensor{\textbf{G}}_{\omega}^{0}$ is the retarded Green's function for free space, provided that both $\textbf{r}$ and $\textbf{r}^{\prime}$ are in the half-space not containing the medium ($z,z^{\prime}<0$). This observation enables us to describe light reflection using the Green's function. However, the incident wave here, generated by a point source, is a spherical wave. To make a connection to the reflectivity tensor, it is essential to construct an object that produces an incident plane wave.

Such an object is, in fact, well known. Consider the following representation of $\tensor{\textbf{G}}_{\omega}^{0}$ \cite{chew1995waves}:
\begin{equation}
\tensor{\textbf{G}}_{\omega}^{0}(\textbf{r},\textbf{r}^{\prime}) = \int \frac{d^{2}\textbf{k}_{\scriptscriptstyle{\parallel}}^{\prime}}{(2\pi)^{2}} \,e^{i\textbf{k}_{\scriptscriptstyle{\parallel}}^{\prime}\cdot (\textbf{r}_{\scriptscriptstyle{\parallel}}-\textbf{r}_{\scriptscriptstyle{\parallel}}^{\prime})}\,\tensor{\boldsymbol{\mathcal{G}}}_{\omega}^{0}(z,z^{\prime};\textbf{k}_{\scriptscriptstyle{\parallel}}^{\prime})
\end{equation}
with
\begin{equation}
\label{freegreen}
\tensor{\boldsymbol{\mathcal{G}}}_{\omega}^{0}(z,z^{\prime};\textbf{k}_{\scriptscriptstyle{\parallel}}^{\prime}) = \frac{ie^{ ik_{z}^{\prime}|z-z^{\prime}|}}{2k_{z}^{\prime}}  \,\tensor{\boldsymbol{\mathcal{P}}}(\textbf{k}_{\mediumsymb{\pm}}^{\prime})
-\frac{c^{2}}{\omega^{2}}\delta (z-z^{\prime})\,\hat{\textbf{z}}\hat{\textbf{z}}.
\end{equation}
Here, $+$ and $-$ are taken for $z>z^{\prime}$ and $z<z^{\prime}$, and $\tensor{\boldsymbol{\mathcal{P}}}(\textbf{k}) \equiv \tensor{\textbf{I}} - k^{\shortminus 2}\textbf{k}\textbf{k}$. Thus, the field generated by the planar source distribution $i\omega \mu_{0} \textbf{J}(\textbf{r}) = e^{i\textbf{k}_{\smallsymb{\parallel}}^{\prime} \cdot \textbf{r}_{\smallsymb{\parallel}}}\delta(z-z^{\prime})\hat{\boldsymbol{\xi}}$ is
\begin{equation}
\label{field_planar}
\int d^{2}\textbf{r}_{\scriptscriptstyle{\parallel}}^{\prime}\,e^{i\textbf{k}_{\smallsymb{\parallel}}^{\prime}\cdot\textbf{r}_{\smallsymb{\parallel}}^{\prime}}\,
\tensor{\textbf{G}}_{\omega}^{0}(\textbf{r},\textbf{r}^{\prime})\cdot\hat{\boldsymbol{\xi}} =
e^{i\textbf{k}_{\smallsymb{\parallel}}^{\prime} \cdot \textbf{r}_{\smallsymb{\parallel}}}\tensor{\boldsymbol{\mathcal{G}}}_{\omega}^{0}(z,z^{\prime};\textbf{k}_{\scriptscriptstyle{\parallel}}^{\prime})\cdot \hat{\boldsymbol{\xi}}.
\end{equation}
This corresponds to a pair of plane waves with the wave vectors $\textbf{k}_{\mediumsymb{\pm}}^{\prime}$ propagating in the two half-spaces $z>z^{\prime}$ and $z<z^{\prime}$.

Now suppose that a scattering medium is placed in one of the half-spaces. We assume that $z^{\prime}<0$ and, as before, that the medium is in the region $z>0$ (e.g., as in Fig. \ref{fig}). Among the two plane waves, only the one in the half-space $z>z^{\prime}$ is reflected from the medium. For $z^{\prime}<z<0$, the right-hand side of Eq. (\ref{field_planar}) reduces to $e^{i\textbf{k}_{\symb{+}}^{\prime}\cdot \textbf{r} }\,\boldsymbol{\mathcal{E}}_{\!\scriptscriptstyle{\perp}}$ with $\boldsymbol{\mathcal{E}}_{\!\scriptscriptstyle{\perp}} = i(2k_{z}^{\prime})^{\shortminus 1}e^{\shortminus ik_{z}^{\prime}z^{\prime}}\tensor{\boldsymbol{\mathcal{P}}}(\textbf{k}_{\mediumsymb{+}}^{\prime}) \cdot \hat{\boldsymbol{\xi}}$. The resulting reflected wave can be either obtained from Eq. (\ref{reflected}) or expressed in terms of the Green's function. Thus, for $z,z^{\prime}<0$,
\begin{equation}
\begin{split}
\label{reflected_rep}
&\frac{ie^{\shortminus ik_{z}^{\prime}z^{\prime}}}{2k_{z}^{\prime}}\int \frac{d^{2}\textbf{k}_{\scriptscriptstyle{\parallel}}}{(2\pi)^{2}} \,e^{i\textbf{k}_{\symb{-}}\cdot \textbf{r}}\,\tensor{\textbf{R}}_{\omega}(\textbf{k}_{\scriptscriptstyle{\parallel}},\textbf{k}_{\scriptscriptstyle{\parallel}}^{\prime})\cdot
\tensor{\boldsymbol{\mathcal{P}}}(\textbf{k}_{\mediumsymb{+}}^{\prime}) \cdot\hat{\boldsymbol{\xi}}\\
&=\int d^{2}\textbf{r}_{\scriptscriptstyle{\parallel}}^{\prime}\,e^{i\textbf{k}_{\smallsymb{\parallel}}^{\prime}\cdot\textbf{r}_{\smallsymb{\parallel}}^{\prime}}
\big[\tensor{\textbf{G}}_{\omega}(\textbf{r},\textbf{r}^{\prime}) - \tensor{\textbf{G}}_{\omega}^{0}(\textbf{r},\textbf{r}^{\prime})\big]\cdot \hat{\boldsymbol{\xi}},
\end{split}
\end{equation}
and $\tensor{\textbf{R}}_{\omega}(\textbf{k}_{\scriptscriptstyle{\parallel}},\textbf{k}_{\scriptscriptstyle{\parallel}}^{\prime})\cdot
\tensor{\boldsymbol{\mathcal{P}}}(\textbf{k}_{\mediumsymb{+}}^{\prime}) = \tensor{\textbf{R}}_{\omega}(\textbf{k}_{\scriptscriptstyle{\parallel}},\textbf{k}_{\scriptscriptstyle{\parallel}}^{\prime})$ due to transversality. As $\hat{\boldsymbol{\xi}}$ is arbitrary, Eq. (\ref{reflected_rep}) without the contraction with $\hat{\boldsymbol{\xi}}$ holds as a tensor identity. Then, the Fourier transform with respect to $\textbf{r}_{\scriptscriptstyle{\parallel}}$ allows us to express the reflectivity tensor in terms of the Green's function as
\begin{equation}
\begin{split}
\label{result}
&\frac{i}{2k_{z}^{\prime}} e^{\shortminus i(k_{z}z + k_{z}^{\prime}z^{\prime})} \,\tensor{\textbf{R}}_{\omega}(\textbf{k}_{\scriptscriptstyle{\parallel}},\textbf{k}_{\scriptscriptstyle{\parallel}}^{\prime})\\
&=\tensor{\boldsymbol{\mathcal{G}}}_{\omega}(z,z^{\prime};\textbf{k}_{\scriptscriptstyle{\parallel}},\textbf{k}_{\scriptscriptstyle{\parallel}}^{\prime}) - \tensor{\boldsymbol{\mathcal{G}}}_{\omega}^{0}(z,z^{\prime};\textbf{k}_{\scriptscriptstyle{\parallel}},\textbf{k}_{\scriptscriptstyle{\parallel}}^{\prime}) \quad \ (z,z^{\prime}<0),
\end{split}
\end{equation}
where
\begin{equation}
\label{fouriergreen}
\tensor{\boldsymbol{\mathcal{G}}}_{\omega}(z,z^{\prime};\textbf{k}_{\scriptscriptstyle{\parallel}},\textbf{k}_{\scriptscriptstyle{\parallel}}^{\prime})\equiv \int d^{2}\textbf{r}_{\scriptscriptstyle{\parallel}}d^{2}\textbf{r}_{\scriptscriptstyle{\parallel}}^{\prime}\, e^{\shortminus i(\textbf{k}_{\smallsymb{\parallel}}\cdot\textbf{r}_{\smallsymb{\parallel}} - \textbf{k}_{\smallsymb{\parallel}}^{\prime}\cdot\textbf{r}_{\smallsymb{\parallel}}^{\prime} )}\,\tensor{\textbf{G}}_{\omega}(\textbf{r},\textbf{r}^{\prime})
\end{equation}
and
\begin{equation}
\label{freegreen2}
\tensor{\boldsymbol{\mathcal{G}}}_{\omega}^{0}(z,z^{\prime};\textbf{k}_{\scriptscriptstyle{\parallel}},\textbf{k}_{\scriptscriptstyle{\parallel}}^{\prime})= (2\pi)^{2}\delta^{(2)}(\textbf{k}_{\scriptscriptstyle{\parallel}}-\textbf{k}_{\scriptscriptstyle{\parallel}}^{\prime})\, \tensor{\boldsymbol{\mathcal{G}}}_{\omega}^{0}(z,z^{\prime};\textbf{k}_{\scriptscriptstyle{\parallel}}^{\prime}).
\end{equation}

As will be shown later, the Onsager symmetry of the electromagnetic response function implies
\begin{equation}
\label{greensymmetry}
\tensor{\boldsymbol{\mathcal{G}}}_{\omega}(z,z^{\prime};\textbf{k}_{\scriptscriptstyle{\parallel}},\textbf{k}_{\scriptscriptstyle{\parallel}}^{\prime})
=\big[\tensor{\boldsymbol{\mathcal{G}}}_{\omega}(z^{\prime},z;-\textbf{k}_{\scriptscriptstyle{\parallel}}^{\prime},-\textbf{k}_{\scriptscriptstyle{\parallel}})\big]^{T}.
\end{equation}
Eqs. (\ref{freegreen}) and (\ref{freegreen2}) show that $\tensor{\boldsymbol{\mathcal{G}}}_{\omega}^{0}$ also satisfies the analogous relation. (This can be viewed as a special case of the above equation.) From Eqs. (\ref{result}) and (\ref{greensymmetry}), the reciprocity relation for light reflection follows:
\begin{equation}
\label{reflection_reciprocity}
\frac{1}{k_{z}^{\prime}}\tensor{\textbf{R}}_{\omega}(\textbf{k}_{\scriptscriptstyle{\parallel}},\textbf{k}_{\scriptscriptstyle{\parallel}}^{\prime}) = \frac{1}{k_{z}}\big[\tensor{\textbf{R}}_{\omega}(-\textbf{k}_{\scriptscriptstyle{\parallel}}^{\prime},-\textbf{k}_{\scriptscriptstyle{\parallel}})\big]^{T},
\end{equation}
which was previously proved under the assumption that spatial dispersion effects are negligible \cite{carminati1998reciprocity}. Eq. (\ref{reflection_reciprocity}) reduces to Eq. (\ref{kerrabsence}) for backscattering ($\textbf{k}_{\scriptscriptstyle{\parallel}}=-\textbf{k}_{\scriptscriptstyle{\parallel}}^{\prime}$) and, hence, implies the absence of Kerr rotation.

As a side note, a similar construction can also be applied to light transmission. If the scattering medium has a finite extent in the $z$ direction, one can define, in a manner analogous to the reflectivity tensor, the transmissivity tensors for right- and left-moving waves (denoted as $\tensor{\textbf{T}}_{\omega}^{\mediumsymb{\pm}}$). It is straightforward to verify that if $z$ and $z^{\prime}$ lie outside of and on opposite sides of the medium,
\begin{equation}
\label{result2}
\frac{i}{2k_{z}^{\prime}} e^{\mediumsymb{\pm} i(k_{z}z - k_{z}^{\prime}z^{\prime})} \,\tensor{\textbf{T}}_{\omega}^{\mediumsymb{\pm}}(\textbf{k}_{\scriptscriptstyle{\parallel}},\textbf{k}_{\scriptscriptstyle{\parallel}}^{\prime})
=\tensor{\boldsymbol{\mathcal{G}}}_{\omega}(z,z^{\prime};\textbf{k}_{\scriptscriptstyle{\parallel}},\textbf{k}_{\scriptscriptstyle{\parallel}}^{\prime}).
\end{equation}
Here, $+$ and $-$ are taken for $z>z^{\prime}$ and $z<z^{\prime}$, and the transversality conditions read $\textbf{k}_{\mediumsymb{\pm}}\cdot \tensor{\textbf{T}}_{\omega}^{\mediumsymb{\pm}}(\textbf{k}_{\scriptscriptstyle{\parallel}},\textbf{k}_{\scriptscriptstyle{\parallel}}^{\prime}) = \tensor{\textbf{T}}_{\omega}^{\mediumsymb{\pm}}(\textbf{k}_{\scriptscriptstyle{\parallel}},\textbf{k}_{\scriptscriptstyle{\parallel}}^{\prime})\cdot \textbf{k}_{\mediumsymb{\pm}}^{\prime}= 0$. Eqs. (\ref{greensymmetry}) and (\ref{result2}) lead to the reciprocity relation for transmission \cite{carminati1998reciprocity}:
\begin{equation}
\label{transmnission_reciprocity}
\frac{1}{k_{z}^{\prime}}\tensor{\textbf{T}}_{\omega}^{\mediumsymb{+}}(\textbf{k}_{\scriptscriptstyle{\parallel}},\textbf{k}_{\scriptscriptstyle{\parallel}}^{\prime}) = \frac{1}{k_{z}}\big[\tensor{\textbf{T}}_{\omega}^{\mediumsymb{-}}(-\textbf{k}_{\scriptscriptstyle{\parallel}}^{\prime},-\textbf{k}_{\scriptscriptstyle{\parallel}})\big]^{T}.
\end{equation}

We now derive Eq. (\ref{greensymmetry}). Consider general linear constitutive relations for time-harmonic fields in nonlocal electrodynamics:
\begin{equation}
\label{constitutive}
\textbf{D} = \widetilde{\boldsymbol{\epsilon}}_{\omega}\textbf{E},\qquad \textbf{H} =\mu_{0}^{\shortminus 1}\textbf{B},
\end{equation}
where the permittivity operator $\widetilde{\boldsymbol{\epsilon}}_{\omega}$ is defined by
\begin{equation}
\label{responsefunc}
\frac{1}{\epsilon_{0}}\big(\widetilde{\boldsymbol{\epsilon}}_{\omega}\textbf{E}\big)(\textbf{r}) \equiv  \textbf{E}(\textbf{r}) + \int d^{3}\textbf{r}^{\prime} \,\tensor{\boldsymbol{\chi}}_{\omega}(\textbf{r},\textbf{r}^{\prime}) \cdot \textbf{E}(\textbf{r}^{\prime}).
\end{equation}
We have adopted the Landau-Lifshitz approach (\S 103 of \cite{LLelectro}), in which the effect of the medium is solely contained in $\widetilde{\boldsymbol{\epsilon}}_{\omega}$, i.e., the permeability is simply taken to be the constant $\mu_0$. The kernel $\tensor{\boldsymbol{\chi}}_{\omega}$ is the electromagnetic response function (nonlocal susceptibility tensor). From the linear response theory, it can be shown that time-reversal symmetry and thermal equilibrium lead to the following Onsager symmetry relation \cite{callen1}:
\begin{equation}
\label{onsagersym}
\tensor{\boldsymbol{\chi}}_{\omega}(\textbf{r},\textbf{r}^{\prime}) = \big[\tensor{\boldsymbol{\chi}}_{\omega}(\textbf{r}^{\prime}\!,\textbf{r})\big]^{T}.
\end{equation}
The dynamics of the fields is given by the macroscopic Maxwell equations:
\begin{equation}
\begin{split}
\label{maxwelleq}
&\nabla\times \textbf{E} = i\omega\textbf{B},\qquad \nabla\cdot\textbf{B} = 0,\\
&\nabla\cdot \textbf{D} = \rho,\qquad \quad \ \ \nabla\times \textbf{H} = \textbf{J}-i\omega\textbf{D}.
\end{split}
\end{equation}
Eqs. (\ref{constitutive}) and (\ref{maxwelleq}) lead to an integro-differential equation for $\textbf{E}$, the electromagnetic wave equation:
\begin{equation}
\label{waveeq}
\widetilde{\mathcal{L}}_{\omega}\textbf{E}\equiv (\omega\mu_{0})^{\shortminus 1} i\!\Laplace_{t}\textbf{E} + i\omega\widetilde{\boldsymbol{\epsilon}}_{\omega}\textbf{E} = \textbf{J},
\end{equation}
where $\Laplace_{t}$ is the transverse Laplacian defined by $\Laplace_{t}\textbf{E} \equiv -\nabla\times(\nabla\times\textbf{E})$. Recall that we assume the scattering medium is entirely in the region $z>0$; that is, $\tensor{\boldsymbol{\chi}}_{\omega}(\textbf{r},\textbf{r}^{\prime}) = 0$ if $z<0$ or $z^{\prime}<0$ \cite{note2}. Hence,
\begin{equation}
\label{vacuumeq}
\big(\widetilde{\mathcal{L}}_{\omega}\textbf{E}\big)(\textbf{r}) = \frac{i}{\omega\mu_{0}}\bigg(\!\Laplace_{t} + \frac{\omega^{2}}{c^{2}}\bigg)\textbf{E}(\textbf{r})\quad
(z<0).
\end{equation}

The Green's function is the kernel associated with the operator $(i\omega\mu_{0})^{\shortminus 1}\widetilde{\mathcal{L}}_{\omega}^{\shortminus 1}$:
\begin{equation}
\label{greenkernel}
\frac{1}{i\omega\mu_{0}}\big(\widetilde{\mathcal{L}}_{\omega}^{\shortminus 1}\textbf{J}\big)(\textbf{r}) \equiv \int d^{3}\textbf{r}^{\prime}\, \tensor{\textbf{G}}_{\omega}(\textbf{r},\textbf{r}^{\prime}) \cdot \textbf{J}(\textbf{r}^{\prime}).
\end{equation}
This definition shows that the symmetry of $\tensor{\textbf{G}}_{\omega}(\textbf{r},\textbf{r}^{\prime})$ is inherited from that of $\widetilde{\mathcal{L}}_{\omega}$. In particular, $\widetilde{\mathcal{L}}_{\omega}$ is a complex symmetric operator, i.e.,
\begin{equation}
\label{complexsym}
\langle\textbf{E}_{1},\widetilde{\mathcal{L}}_{\omega}\textbf{E}_{2}\rangle = \langle\widetilde{\mathcal{L}}_{\omega}\textbf{E}_{1},\textbf{E}_{2}\rangle,
\end{equation}
for $\textbf{E}_{1}$ and $\textbf{E}_{2}$ belonging to a suitable class of vector-valued functions. Here, we define
\begin{equation}
\langle\textbf{u},\textbf{v}\rangle \equiv \int\! d^{3}\textbf{r} \,\textbf{u}(\textbf{r}) \cdot \textbf{v}(\textbf{r}).
\label{innerprod}
\end{equation}
Eq. (\ref{complexsym}) follows from Eq. (\ref{onsagersym}) and $\langle\textbf{E}_{1},\Laplace_{t}\textbf{E}_{2}\rangle = \langle\Laplace_{t}\textbf{E}_{1},\textbf{E}_{2}\rangle$ (integration by parts without a surface term). Defining $\textbf{J}_{1}\equiv \widetilde{\mathcal{L}}_{\omega}\textbf{E}_{1}$ and $\textbf{J}_{2}\equiv \widetilde{\mathcal{L}}_{\omega}\textbf{E}_{2}$, we see that $\widetilde{\mathcal{L}}_{\omega}^{\shortminus 1}$ is also complex symmetric:
\begin{equation}
\langle\widetilde{\mathcal{L}}_{\omega}^{\shortminus 1}\textbf{J}_{1},\textbf{J}_{2}\rangle = \langle\textbf{J}_{1},\widetilde{\mathcal{L}}_{\omega}^{\shortminus 1}\textbf{J}_{2}\rangle.
\end{equation}
Together with Eq. (\ref{greenkernel}), this implies the symmetry relation
\begin{equation}
\tensor{\textbf{G}}_{\omega}(\textbf{r},\textbf{r}^{\prime}) = \big[\tensor{\textbf{G}}_{\omega}(\textbf{r}^{\prime},\textbf{r})\big]^{T},
\end{equation}
and Eq. (\ref{greensymmetry}) follows by virtue of Eq. (\ref{fouriergreen}).

In the above analysis, one should ensure that $\tensor{\textbf{G}}_{\omega}$ or, equivalently, $\widetilde{\mathcal{L}}_{\omega}^{\shortminus 1}$ is uniquely determined. Notice that $\tensor{\textbf{G}}_{\omega}$ is the retarded Green's function. In other words, $\widetilde{\mathcal{L}}_{\omega}^{\shortminus 1}\textbf{J}$ is an outgoing wave (with the effect of scattering included), meaning that no radiation originates from infinity. [Only then can the right-hand side of Eq. (\ref{reflected_rep}) be identified as a pure reflected wave.] Intuitively, we expect that restricting $\tensor{\textbf{G}}_{\omega}$ to be the retarded Green's function guarantees its uniqueness; a given current distribution should result in a unique outgoing electromagnetic field distribution.

It is possible to make the uniqueness argument more formal. We sketch the arguments here; a more thorough mathematical treatment is given in the \hyperlink{supp}{Supplemental} \hyperlink{supp}{Material}. As is often done to ensure causality, we add an arbitrarily small positive imaginary part to the frequency ($\omega\,\to\,\omega_{\symb{+}} \equiv \omega + i\eta$) \cite{note3}. The frequency shift $i\eta$ makes the entire space slightly dissipative. This fact is readily seen for a homogeneous, lossless, propagating medium. Consider a plane wave proportional to $e^{i\textbf{k}\cdot\textbf{r}}$ and a dispersion relation $\omega = f(\textbf{k})$. Upon shifting $\omega$ by $i\eta$, $\textbf{k}$ should change by an amount $\delta\textbf{k}$ satisfying $\delta\textbf{k} \cdot (\partial f/\partial \textbf{k}) = i\eta$ to maintain the dispersion relation. The additional factor $e^{i\,\delta\textbf{k}\cdot\textbf{r}}$ exponentially decays along the direction of propagation (parallel to $\partial f/\partial \textbf{k}$).

In the presence of the dissipation introduced by $i\eta$, a wave $\textbf{E}(\textbf{r})$ originating from infinitely far away must have a divergent amplitude there. Otherwise it vanishes at any finite $\textbf{r}$, i.e., after having traveled and been dissipated over an infinite distance. For the same reason, we expect that a wave is exponentially suppressed at infinity if and only if it is outgoing. Two such waves $\textbf{E}_{1}$ and $\textbf{E}_{2}$ satisfy $\langle\textbf{E}_{1},\Laplace_{t}\textbf{E}_{2}\rangle = \langle\Laplace_{t}\textbf{E}_{1},\textbf{E}_{2}\rangle$ (i.e., no surface term arises) and, hence, Eq. (\ref{complexsym}). Moreover, one can select  outgoing waves by demanding $\textbf{E}$ to be square integrable. This requirement, in particular, excludes nontrivial solutions of the homogeneous equation $\widetilde{\mathcal{L}}_{\omega_{\smallsymb{+}}}\textbf{E}=0$ because they correspond to waves generated at infinity. It then follows that the solution of the inhomogeneous equation $\widetilde{\mathcal{L}}_{\omega_{\smallsymb{+}}}\textbf{E}=\textbf{J}$ is unique.

In summary, we have shown, in the framework of nonlocal electrodynamics, that the Onsager symmetry of the electromagnetic response function implies the absence of Kerr rotation in backscattering and, more generally, the principle of optical reciprocity. An important observation is that the symmetry property of the response function is inherited by the Green's function and then by the reflectivity tensor.

\begin{acknowledgments}
We thank Peter Armitage, Andrey Chubukov, Alexander Fried, Pavan Hosur, Aharon Kapitulnik, Mark H. Kim,  Robert Laughlin, Ga In Lee, Dirk van der Marel, Joseph Orenstein,  Srinivas Raghu, and Victor Yakovenko  for helpful discussions.
This work was supported in part by the NSF under Grant No. DMR-1265593.
\end{acknowledgments}

\appendix*
\section{Supplemental Material}
\hypertarget{supp}{}
\setcounter{equation}{0}

In the main body of the paper, a symmetry relation for the Green's function [Eq. (\ref{greensymmetry})] played a crucial role in proving the absence of the polar Kerr effect. To establish this relation, we argued, first, that a unique outgoing wave solution exists for a given source distribution and, second, that $\langle\textbf{E}_{1},\Laplace_{t}\textbf{E}_{2}\rangle = \langle\Laplace_{t}\textbf{E}_{1},\textbf{E}_{2}\rangle$ for two outgoing waves $\textbf{E}_{1}$ and $\textbf{E}_{2}$. [Recall that $\langle\textbf{u},\textbf{v}\rangle \equiv \int\! d^{3}\textbf{r} \,\textbf{u}(\textbf{r}) \cdot \textbf{v}(\textbf{r})$.] We formally prove these statements here. As before, the frequency is assumed to have an arbitrarily small positive imaginary part, i.e., is of the form $\omega_{\symb{+}} = \omega + i\eta$.

To proceed further, a precise definition of an outgoing wave must be given. In the main text, we have examined wave propagation in a homogeneous, lossless medium. Based on this, it was argued that shifting the frequency by $i\eta$ leads to dissipation, enabling us to select outgoing waves by considering only square-integrable field configurations. More generally, when the medium may be inhomogeneous, it is unclear how to explicitly describe wave propagation. To account for such cases, we take square integrability as the defining property of an outgoing wave. That is, given that the entire space is dissipative, we restrict to field configurations belonging to the Hilbert space
\begin{equation}
L^{2} \equiv \{\textbf{E}: \|\textbf{E}\| <\infty\} \quad \mbox{$\big(\|\textbf{E}\| \equiv \sqrt{\langle\textbf{E}^{\ast},\textbf{E}\rangle}\big)$}.
\end{equation}

In this setting, it is important to ensure that the frequency shift $\omega \to \omega_{\symb{+}} = \omega + i\eta$ actually introduces dissipation. We have provided evidence for this but only in the case of homogeneous media. In general, one should demonstrate that the dissipated power \cite{LLelectro} given by
\begin{equation}
\label{powerloss}
P_{\omega_{\smallsymb{+}}}[\textbf{E}] \equiv  \mathrm{Im}\Big(\langle\textbf{E}^{\ast},\omega_{\symb{+}}\widetilde{\boldsymbol{\epsilon}}_{\omega_{\smallsymb{+}}}\textbf{E}\rangle+\langle\textbf{H}_{\omega_{\smallsymb{+}}}^{\ast},
\omega_{\symb{+}}\mu_{0}\textbf{H}_{\omega_{\smallsymb{+}}}\rangle\Big)
\end{equation}
is positive, where $\textbf{H}_{\omega_{\smallsymb{+}}} \equiv -i\big(\omega_{\symb{+}}\mu_{0}\big)^{\shortminus 1} \nabla\times\textbf{E}$. The positivity condition to be verified is that
\begin{equation}
\label{positivity}
P_{\omega_{\smallsymb{+}}}[\textbf{E}] \ge \lambda \|\textbf{E}\|^{2}
\end{equation}
for some $\lambda>0$. This is different from $P_{\omega_{\smallsymb{+}}}[\textbf{E}]>0$, which does not rule out the possibility that $P_{\omega_{\smallsymb{+}}}[\textbf{E}]$ vanishes for an ``almost'' square-integrable field configuration, e.g., a plane wave. (In such a case, $P_{\omega_{\smallsymb{+}}}[\textbf{E}]$ can be  arbitrarily close to zero for a fixed $\|\textbf{E}\|$.) Notice that
\begin{equation}
\label{powerloss2}
P_{\omega_{\smallsymb{+}}}[\textbf{E}] = P_{\omega}[\textbf{E}] + 2\eta\, U_{\omega}[\textbf{E}],
\end{equation}
where
\begin{equation}
\label{field_energy}
U_{\omega}[\textbf{E}] \equiv \frac{1}{2} \mathrm{Re}\!\left[\bigg\langle\textbf{E}^{\ast},\!\frac{d(\omega\widetilde{\boldsymbol{\epsilon}}_{\omega})}{d\omega}\textbf{E}\bigg\rangle + \big\langle\textbf{H}_{\omega}^{\ast},\mu_{0}\textbf{H}_{\omega}\!\big\rangle\right].
\end{equation}
The second law of thermodynamics requires that $P_{\omega}[\textbf{E}]\ge 0$ \cite{note4}, where the equality holds in the absence of dissipation (e.g., if the medium is lossless or $\textbf{E}$ is localized outside the medium). When $P_{\omega}[\textbf{E}]/\|\textbf{E}\|^{2}\to 0$, $U_{\omega}[\textbf{E}]$ can be interpreted as the field energy (see Section 2.3 of \cite{agranovich2013crystal}), which must be positive, i.e., $U_{\omega}[\textbf{E}]\ge\kappa\|\textbf{E}\|^{2}$ for some $\kappa>0$. From the behaviors of $P_{\omega}[\textbf{E}]$ and $U_{\omega}[\textbf{E}]$, we see that Eq. (\ref{positivity}) with $\lambda=O(\eta)$ holds.

We now prove that the solution of the equation
\begin{equation}
\widetilde{\mathcal{L}}_{\omega_{\smallsymb{+}}}\textbf{E}\equiv (\omega_{\symb{+}}\mu_{0})^{\shortminus 1} i\!\Laplace_{t}\textbf{E} + i\omega_{\symb{+}}\widetilde{\boldsymbol{\epsilon}}_{\omega_{\smallsymb{+}}}\textbf{E} = \textbf{J}
\end{equation}
is unique if it exists. Assuming (as we will show later to be the case) that there is no surface term, integration by parts gives $\langle\textbf{E}^{\ast},-\!\Laplace_{t}\textbf{E}\rangle=\langle\nabla\times\textbf{E}^{\ast},\nabla\times\textbf{E}\rangle$ and hence
\begin{equation}
\label{positivity2}
-\mathrm{Re}\langle\textbf{E}^{\ast},\widetilde{\mathcal{L}}_{\omega_{\smallsymb{+}}}\textbf{E}\rangle = P_{\omega_{\smallsymb{+}}}[\textbf{E}]\ge \lambda \|\textbf{E}\|^{2},
\end{equation}
where the inequality is simply Eq. (\ref{positivity}).
Therefore, $\widetilde{\mathcal{L}}_{\omega_{\smallsymb{+}}}\textbf{E}=0$ implies $\textbf{E}=0$. Suppose that there exist two solutions $\textbf{E}_{1}$ and $\textbf{E}_{2}$ for a given source distribution $\textbf{J}$, i.e., $\textbf{J}=\widetilde{\mathcal{L}}_{\omega_{\smallsymb{+}}}\textbf{E}_{1} = \widetilde{\mathcal{L}}_{\omega_{\smallsymb{+}}}\textbf{E}_{2}$. Then we must have $\textbf{E}_{1} = \textbf{E}_{2}$ because $\widetilde{\mathcal{L}}_{\omega_{\smallsymb{+}}}(\textbf{E}_{1} -\textbf{E}_{2}) = 0$.

It remains to show that outgoing wave solutions exist and that one can integrate by parts without surface terms in Eq. (\ref{complexsym}) of the main text and in Eq. (\ref{positivity2}). More precisely, we aim to establish the following: first, that the equation $\widetilde{\mathcal{L}}_{\omega_{\smallsymb{+}}}\textbf{E}=\textbf{J}$ has solutions for all $\textbf{J}$ in $L^{2}$; second, that each solution $\widetilde{\mathcal{L}}_{\omega_{\smallsymb{+}}}^{\shortminus 1}\textbf{J}$ belongs to a subset of $L^{2}$ for which the integration by parts formulas we have used are valid. Both statements refer to the properties of $\widetilde{\mathcal{L}}_{\omega_{\smallsymb{+}}}$ as a linear operator on $L^{2}$. Because $\widetilde{\mathcal{L}}_{\omega_{\smallsymb{+}}}$ is a linear combination of $\widetilde{\boldsymbol{\epsilon}}_{\omega_{\smallsymb{+}}}$ and $\Laplace_{t}$, we begin by considering them separately.

Concerning the permittivity operator $\widetilde{\boldsymbol{\epsilon}}_{\omega_{\smallsymb{+}}}$, we make an additional assumption for further progress: we require that it is a bounded operator on $L^{2}$, i.e.,
\begin{equation}
\label{epsilonbound}
\|\widetilde{\boldsymbol{\epsilon}}_{\omega_{\smallsymb{+}}}\textbf{E}\| \le \gamma \|\textbf{E}\|
\end{equation}
for some $\gamma>0$. This condition can be derived from a reasonable assumption on the electromagnetic response function; it is sufficient if there exists a positive function $M(\textbf{r},\textbf{r}^{\prime})$ satisfying
\begin{equation}
\label{bound1}
|\tensor{\boldsymbol{\chi}}_{\omega_{\smallsymb{+}}}(\textbf{r},\textbf{r}^{\prime})\cdot \textbf{s}| \le M(\textbf{r},\textbf{r}^{\prime}) |\textbf{s}|
\end{equation}
for all $\textbf{r}$, $\textbf{r}^{\prime},\textbf{s}$ in $\mathbb{R}^{3}$ and
\begin{equation}
\label{bound2}
\int d^{3}\textbf{r}^{\prime}\, M(\textbf{r},\textbf{r}^{\prime}) \le \beta,\quad \int d^{3}\textbf{r}\, M(\textbf{r},\textbf{r}^{\prime}) \le \beta^{\prime}
\end{equation}
for some $\beta, \beta^{\prime}>0$. Eqs. (\ref{bound1}) and (\ref{bound2}) imply $\| \epsilon_{0}^{\shortminus 1}\widetilde{\boldsymbol{\epsilon}}_{\omega_{\smallsymb{+}}}\textbf{E} - \textbf{E}\| \le (\beta\beta^{\prime})^{1/2}\|\textbf{E}\|$ due to Schur test (Theorem 5.2 of \cite{halmos1978bounded}). Then, the triangle inequality, i.e., $\|\textbf{u}+\textbf{v}\| \le \|\textbf{u}\|+\|\textbf{v}\|$ for all $\textbf{u}, \textbf{v} \in L^{2}$, leads to  Eq. (\ref{epsilonbound}) with $\gamma=\epsilon_{0} [(\beta\beta^{\prime})^{1/2}+1]$.

The transverse Laplacian $\Laplace_{t}$ \cite{note5} is an unbounded operator. The domain of $\Laplace_{t}$ is $\mathcal{D}(\Laplace_{t}) \equiv \big\{\textbf{E}\in L^{2}:\Laplace_{t}\textbf{E} \in L^{2}\big\}$, a dense subset of $L^{2}$. This is in marked contrast to $\widetilde{\boldsymbol{\epsilon}}_{\omega_{\smallsymb{+}}}$, whose domain defined analogously coincides with the entire $L^{2}$ because of Eq. (\ref{epsilonbound}). A notable property of $\Laplace_{t}$ is that it is a closed operator, i.e., if two sequences $\{\textbf{E}_{n}\in\mathcal{D}(\Laplace_{t})\}$ and $\{\Laplace_{t}\textbf{E}_{n}\}$ converge respectively to $\textbf{E}$ and $\textbf{F}$ both in $L^{2}$, then $\textbf{E}$ belongs to $\mathcal{D}(\Laplace_{t})$ and $\textbf{F}=\Laplace_{t}\textbf{E}$. [Due to Eq. (\ref{epsilonbound}), $\{\widetilde{\boldsymbol{\epsilon}}_{\omega_{\smallsymb{+}}}\textbf{E}_{n}\}$ converges to $\widetilde{\boldsymbol{\epsilon}}_{\omega_{\smallsymb{+}}}\textbf{E}$ if $\{\textbf{E}_{n}\}$ converges to $\textbf{E}$, so $\widetilde{\boldsymbol{\epsilon}}_{\omega_{\smallsymb{+}}}$ is automatically a closed operator.] Furthermore, $\Laplace_{t}$ is self-adjoint, i.e., $\mathcal{D}(\Laplace_{t}^{\dagger})=\mathcal{D}(\Laplace_{t})$ and $\Laplace_{t}^{\dagger} = \Laplace_{t}$ \cite{richtmyer1981principles}.

Next, consider $\widetilde{\mathcal{L}}_{\omega_{\smallsymb{+}}} = A\!\Laplace_{t} + B\widetilde{\boldsymbol{\epsilon}}_{\omega_{\smallsymb{+}}}$, where $iA^{\shortminus 1}\equiv \omega_{\symb{+}}\mu_{0}$ and $B=i\omega_{\symb{+}}$, acting on $\mathcal{D}(\widetilde{\mathcal{L}}_{\omega_{\smallsymb{+}}}) \equiv \big\{\textbf{E}\in L^{2}:\widetilde{\mathcal{L}}_{\omega_{\smallsymb{+}}} \textbf{E} \in L^{2}\big\}$. The triangle inequality and Eq. (\ref{epsilonbound}) give
\begin{equation}
\begin{split}
&\|\widetilde{\mathcal{L}}_{\omega_{\smallsymb{+}}}\textbf{E}\|\le |A|\|\!\Laplace_{t}\textbf{E}\| + \gamma|B|\|\textbf{E}\|,\\
&|A|\|\!\Laplace_{t}\textbf{E}\|\le \|\widetilde{\mathcal{L}}_{\omega_{\smallsymb{+}}}\textbf{E}\| + \gamma |B| \|\textbf{E}\|,
\end{split}
\end{equation}
which, respectively, imply that $\mathcal{D}(\Laplace_{t})\subset\mathcal{D}(\widetilde{\mathcal{L}}_{\omega_{\smallsymb{+}}})$ and that $\mathcal{D}(\widetilde{\mathcal{L}}_{\omega_{\smallsymb{+}}})\subset\mathcal{D}(\Laplace_{t})$. Thus, $\mathcal{D}(\widetilde{\mathcal{L}}_{\omega_{\smallsymb{+}}})=\mathcal{D}(\Laplace_{t})$, and using the definition of a closed operator, it is straightforward to check that $\widetilde{\mathcal{L}}_{\omega_{\smallsymb{+}}}$ is closed.

The adjoint of $\widetilde{\mathcal{L}}_{\omega_{\smallsymb{+}}}$ is defined as follows: $\tilde {\textbf{E} }$ is an element of $\mathcal{D}(\widetilde{\mathcal{L}}_{\omega_{\smallsymb{+}}}^{\dagger})$ if and only if $\langle\tilde{\textbf{E}}^{\ast},\widetilde{\mathcal{L}}_{\omega_{\smallsymb{+}}}\textbf{E}\rangle = \langle\tilde{\textbf{J}}^{\ast},\textbf{E}\rangle$ for a $\tilde{\textbf{J}}$ in $L^{2}$ and for all $\textbf{E}$ in $\mathcal{D}(\widetilde{\mathcal{L}}_{\omega_{\smallsymb{+}}})$, in which case we set $\tilde{\textbf{J}} =\widetilde{\mathcal{L}}_{\omega_{\smallsymb{+}}}^{\dagger}\tilde{\textbf{E}}$. From the fact that all bounded operators have bounded adjoints \cite{richtmyer1981principles} and that $\Laplace_{t}$ is self-adjoint, one can show that $\mathcal{D}(\widetilde{\mathcal{L}}_{\omega_{\smallsymb{+}}}^{\dagger}) = \mathcal{D}(\Laplace_{t}) = \mathcal{D}(\widetilde{\mathcal{L}}_{\omega_{\smallsymb{+}}})$.

We now prove that the equation $\textbf{J}=\widetilde{\mathcal{L}}_{\omega_{\smallsymb{+}}}\textbf{E}$ has a solution for each $\textbf{J}$ in $L^{2}$. That is, the range of $\widetilde{\mathcal{L}}_{\omega_{\smallsymb{+}}}$ (or, equivalently, the domain of $\widetilde{\mathcal{L}}_{\omega_{\smallsymb{+}}}^{\shortminus 1}$) defined by
\begin{equation}
\mathcal{R}(\widetilde{\mathcal{L}}_{\omega_{\smallsymb{+}}}) \equiv \big\{\textbf{J}\in L^{2}:\textbf{J}=\widetilde{\mathcal{L}}_{\omega_{\smallsymb{+}}}\textbf{E},\, \textbf{E}\in \mathcal{D}(\widetilde{\mathcal{L}}_{\omega_{\smallsymb{+}}})\big\},
\end{equation}
coincides with $L^{2}$.
To see this,
we first show that there is no nonzero element of $L^{2}$ orthogonal to $\mathcal{R}(\widetilde{\mathcal{L}}_{\omega_{\smallsymb{+}}})$. If such an element $\textbf{E}_{\perp}$ existed, $\langle\textbf{E}_{\perp}^{\ast},\widetilde{\mathcal{L}}_{\omega_{\smallsymb{+}}}\textbf{E}\rangle=0$ for all $\textbf{E}$ in $\mathcal{D}(\widetilde{\mathcal{L}}_{\omega_{\smallsymb{+}}})$. It would then follow that $\textbf{E}_{\perp}$ belongs to $\mathcal{D}(\widetilde{\mathcal{L}}_{\omega_{\smallsymb{+}}}^{\dagger})=\mathcal{D}(\widetilde{\mathcal{L}}_{\omega_{\smallsymb{+}}})$ and that $\widetilde{\mathcal{L}}_{\omega_{\smallsymb{+}}}^{\dagger}\textbf{E}_{\perp}=0$. Consequently, $\langle\textbf{E}_{\perp}^{\ast},\widetilde{\mathcal{L}}_{\omega_{\smallsymb{+}}}\textbf{E}_{\perp}\rangle =
\langle\textbf{E}_{\perp}^{\ast},\widetilde{\mathcal{L}}_{\omega_{\smallsymb{+}}}^{\dagger}\textbf{E}_{\perp}\rangle^{\ast} = 0$,
contradicting Eq. (\ref{positivity2}). Therefore, $\mathcal{R}(\widetilde{\mathcal{L}}_{\omega_{\smallsymb{+}}})$ has a trivial orthogonal complement and hence is dense in $L^{2}$. Furthermore, Eq. (\ref{positivity2}) and Schwarz's inequality lead to
\begin{equation}
\begin{split}
\|\widetilde{\mathcal{L}}_{\omega_{\smallsymb{+}}}\textbf{E}\| \, \|\textbf{E}\| &\ge\big|\langle\textbf{E}^{\ast},\widetilde{\mathcal{L}}_{\omega_{\smallsymb{+}}}\textbf{E}\rangle\big|\ge\big|\mathrm{Re}\langle\textbf{E}^{\ast},\widetilde{\mathcal{L}}_{\omega_{\smallsymb{+}}}\textbf{E}\rangle\big|\\
&\ge \lambda\|\textbf{E}\|^{2},
\end{split}
\end{equation}
whence
\begin{equation}
\label{estimation}
\|\textbf{E}=\widetilde{\mathcal{L}}_{\omega_{\smallsymb{+}}}^{\shortminus 1}\textbf{J}\| \le \lambda^{\shortminus 1}\|\textbf{J}\|.
\end{equation}
Eq. (\ref{estimation}) implies that given a sequence $\{\textbf{J}_{n}\in\mathcal{R}(\widetilde{\mathcal{L}}_{\omega_{\smallsymb{+}}})\}$ converging in $L^{2}$, $\{\textbf{E}_{n}\equiv\widetilde{\mathcal{L}}_{\omega_{\smallsymb{+}}}^{\shortminus 1}\textbf{J}_{n}\in \mathcal{D}(\widetilde{\mathcal{L}}_{\omega_{\smallsymb{+}}})\}$ is also convergent. Let $\textbf{J}_{\infty}$ and $\textbf{E}_{\infty}$ be the limits of these sequences. Because $\widetilde{\mathcal{L}}_{\omega_{\smallsymb{+}}}$ is a closed operator, $\textbf{E}_{\infty}$ belongs to $\mathcal{D}(\widetilde{\mathcal{L}}_{\omega_{\smallsymb{+}}})$, and $\textbf{J}_{\infty}=\widetilde{\mathcal{L}}_{\omega_{\smallsymb{+}}}\textbf{E}_{\infty}$. Therefore, $\textbf{J}_{\infty}$ is an element of $\mathcal{R}(\widetilde{\mathcal{L}}_{\omega_{\smallsymb{+}}})$, i.e., $\mathcal{R}(\widetilde{\mathcal{L}}_{\omega_{\smallsymb{+}}})$ is a closed subset of $L^{2}$. As it is dense and closed, $\mathcal{R}(\widetilde{\mathcal{L}}_{\omega_{\smallsymb{+}}})=L^{2}$.

Knowing $\mathcal{D}(\widetilde{\mathcal{L}}_{\omega_{\smallsymb{+}}}) = \mathcal{D}(\widetilde{\mathcal{L}}_{\omega_{\smallsymb{+}}}^{\dagger}) = \mathcal{D}(\Laplace_{t})$, we can justify why it was possible to integrate by parts without surface terms in Eq. (\ref{complexsym}) of the main text and in Eq. (\ref{positivity2}). It suffices to prove that
\begin{equation}
\label{byparts}
\langle\textbf{E}_{1},\Laplace_{t}\textbf{E}_{2}\rangle = - \langle\nabla\times\textbf{E}_{1},\nabla\times\textbf{E}_{2}\rangle = \langle\Laplace_{t}\textbf{E}_{1},\textbf{E}_{2}\rangle
\end{equation}
for all $\textbf{E}_{1},\textbf{E}_{2}\in\mathcal{D}(\Laplace_{t})$. According to Parseval-Plancherel theorem, the Fourier transform defined by
\begin{equation}
(\mathcal{F}\textbf{E})(\textbf{k}) \equiv \frac{1}{(2\pi)^{3/2}}\int d^{3}\textbf{r}\, \textbf{E}(\textbf{r}) \,e^{\shortminus i\textbf{k}\cdot\textbf{r}}
\end{equation}
is a unitary map of $L^{2}$ onto itself. With $\textbf{V}\equiv\textbf{E}-\Laplace_{t}\textbf{E}$, we have
\begin{equation}
\begin{split}
\label{curlbound}
\|\nabla\times\textbf{E}\| &= \|\textbf{k}\times\mathcal{F}\textbf{E}\| =\|(1+k^{2})^{\shortminus 1}\textbf{k}\times\mathcal{F}\textbf{V}\|\\
&\le \|(1+k^{2})^{\shortminus 1}k\mathcal{F}\textbf{V}\|\le 2^{\shortminus 1}\|\mathcal{F}\textbf{V}\|\\
&=2^{\shortminus 1}\|\textbf{V}\|\le2^{\shortminus 1}(\|\textbf{E}\| + \|\!\Laplace_{t}\textbf{E}\|).
\end{split}
\end{equation}
This inequality shows that if both $\textbf{E}$ and $\Laplace_{t}\textbf{E}$ belong to $L^{2}$, then so does $\nabla\times\textbf{E}$. Hence, all six quantities entering $\langle\cdot\,,\cdot\rangle$ in Eq. (\ref{byparts}) are elements of $L^{2}$. It also follows from Parseval-Plancherel theorem that $\langle\textbf{u},\textbf{v}\rangle = \langle(\mathcal{F}\textbf{u}^{\ast})^{\ast},\mathcal{F}\textbf{v}\rangle$ for all $\textbf{u},\textbf{v}\in L^{2}$. Then, it is straightforward to verify Eq. (\ref{byparts}) by considering the Fourier transforms of the quantities in $\langle\cdot\,,\cdot\rangle$.

Thus far, we have formally proved certain properties of outgoing waves leading to the desired symmetry of the Green's function [Eq. (\ref{greensymmetry}) of the main text]. To be completely rigorous, however, one should ensure the existence of the limit $\eta \to 0^{\scriptscriptstyle{+}}$, called the \emph{limiting absorption principle} \cite{agmon1975spectral,debievre1991spectral}. Extrapolation of our result does not work because, e.g., $\lambda^{\shortminus 1} =O(\eta^{\shortminus 1})$ in Eq. (\ref{estimation}) diverges in this limit. To the best of our knowledge, there is no result sufficiently broad in scope to establish the limiting absorption principle for the nonlocal electrodynamics problem we have considered.


\begin{thebibliography}{9}
\bibitem{ignore_nonlocal} The wavelength of light in Refs. \cite{xia_srru, 14schemm_science, 15schemm_prb, xia_ybco, he11science, karapetyan12prl, karapetyan14prl} is on the order of 1 $\mathrm{\mu m}$, which is much longer than the optical response range in the materials studied. Ignoring spatial dispersion is, typically, a good approximation in such a case.
\bibitem{note0} Natural optical activity is characterized by the rank-3 tensor $\gamma_{ijl}$ in the constitutive relation $D_{i} = \epsilon_{ij} E_{j} + \gamma_{ijl} \partial_{l} E_{j}$. The tensor $\gamma_{ijl}$ is nonzero only if inversion symmetry is broken.
\bibitem{note2} Notice that $z=0$ does not need to be a physical boundary. The medium should be well within the region $z>0$ so that its optical response has no impact on $z<0$.
\bibitem{note4} We limit our discussion to passive media, which do not spontaneously generate energy.
\bibitem{note5} All derivatives should be understood as distributional derivatives.
\end{thebibliography}
\end{document}